\documentclass[aps,prl,reprint, superscriptaddress]{revtex4-1}

\usepackage{amsmath,amssymb, amsthm,amstext}
\usepackage{natbib}
\usepackage{graphicx}
\usepackage{color}
\usepackage{array, enumerate}
\usepackage{bm}
\usepackage{multirow}
\usepackage[breaklinks,colorlinks,citecolor=blue]{hyperref}
\usepackage{braket}

\def\be{\begin{equation}}
\def\ee{\end{equation}}
\newcommand{\Tm}{T_{\rm melt}}

\newcommand{\xim}{\bm{\xi}_m}

\begin{document}
\title{Probing Crust Meltdown in Inspiraling Binary Neutron Stars}

\author{Zhen Pan}
\email{zpan@perimeterinstitute.ca}
\affiliation{Perimeter Institute for Theoretical Physics, Waterloo, Ontario N2L 2Y5, Canada}
\author{Zhenwei Lyu}
\affiliation{Perimeter Institute for Theoretical Physics, Waterloo, Ontario N2L 2Y5, Canada}
\affiliation{University of Guelph, Guelph, Ontario N2L 3G1, Canada}
\author{B\'{e}atrice Bonga}
\affiliation{Institute for Mathematics, Astrophysics and Particle Physics, Radboud University, 6525 AJ Nijmegen, The Netherlands}
\affiliation{Perimeter Institute for Theoretical Physics, Waterloo, Ontario N2L 2Y5, Canada}
\author{N\'{e}stor Ortiz}
\affiliation{Instituto de Ciencias Nucleares, Universidad Nacional Aut\'onoma de M\'exico,	Circuito Exterior C.U., A.P. 70-543, M\'exico D.F. 04510, M\'exico}
\author{Huan Yang}
\email{hyang@perimeterinstitute.ca}
\affiliation{Perimeter Institute for Theoretical Physics, Waterloo, Ontario N2L 2Y5, Canada}
\affiliation{University of Guelph, Guelph, Ontario N2L 3G1, Canada}

\begin{abstract}
  Thanks to recent measurements of
  tidal deformability and  radius,  the nuclear equation of state and structure of neutron stars are  now better understood.
  Here, we show that through resonant tidal excitations in a binary inspiral, the neutron crust generically undergoes elastic-to-plastic transition, which leads to crust heating and eventually meltdown. This process could induce $\sim \mathcal{O}(0.1)$ phase shift in the gravitational waveform.
  Detecting the timing and induced phase shift of this crust meltdown will shed light on the crust structure, such as the core-crust transition density, which previous measurements are insensitive to. A direct search using GW170817 data has not found this signal, possibly due to limited signal-to-noise ratio. We predict that such signal may be observable with Advanced LIGO Plus and more likely with third-generation gravitational-wave detectors such as the Einstein Telescope and Cosmic Explorer.
\end{abstract}
\maketitle

\noindent{\bf Introduction.}
Inspiraling neutron stars deform under  mutual tidal interactions. In the adiabatic limit, the star's induced quadrupole moment is directly proportional to the tidal gravitational field, with the proportionality constant given by the tidal Love number. Deformed neutron stars orbit  each other differently from black holes with the same masses, and the phase difference can be used to measure the tidal Love number \cite{Flanagan:2007ix}, as shown in the analysis of GW170817 \cite{Abbott:2018exr}. Together with neutron star radius measurements \cite{Miller:2019cac}, maximum mass estimates \cite{Rezzolla:2017aly} and possibly post-merger electromagnetic signals \cite{Radice:2017lry}, the star's equation of state (EoS) is now better constrained.

In addition to adiabatic tides, tidal interaction can excite internal modes of neutron stars as the binary sweeps through the inspiral frequency range.
The pressure (p-) and fundamental (f-)modes \cite{Kokkotas1999} will not be fully excited as their frequencies are generally higher than the inspiral frequency, although it has been suggested that early excitation of f-modes may be observed in the late inspiral stage \cite{Pratten:2019sed}. Gravity modes may be fully excited, but their couplings to tidal gravitational fields are so small that the induced phase shifts are $\mathcal{O}(10^{-3})$ or smaller \cite{Lai1994,Yu:2016ltf}.
Resonance of rotational modes has also been investigated assuming a  rotational frequency of a few $\times 10^2$ Hz \cite{Ho:1998hq,Lai:2006pr,Flanagan:2006sb,Poisson2020}, whereas the fastest rotating  pulsar known in a binary neutron star  system has a frequency of $\sim 60$ Hz \cite{Burgay:2003jj,Andrews2019}.

The interface (i-)modes \cite{McDermott1985,McDermott1988}, excited at the interface of the fluid core and solid crust, have frequencies around several tens to a few hundred Hertz, depending on the star's equation of state and prescription of the crust. The resonance of i-modes was proposed to explain precursors of short gamma-ray bursts due to possible crust failures \cite{Tsang2012}.
We observe that through excitation of i-modes, the crustal material actually reaches its elastic limit well before the mode resonance. After reaching this threshold the crust undergoes an elastic-to-plastic transition and the tidal driving starts to heat up the crust. The whole process ends with the meltdown of the crust in tens of cycles.

\begin{figure*}
\includegraphics[scale=0.42]{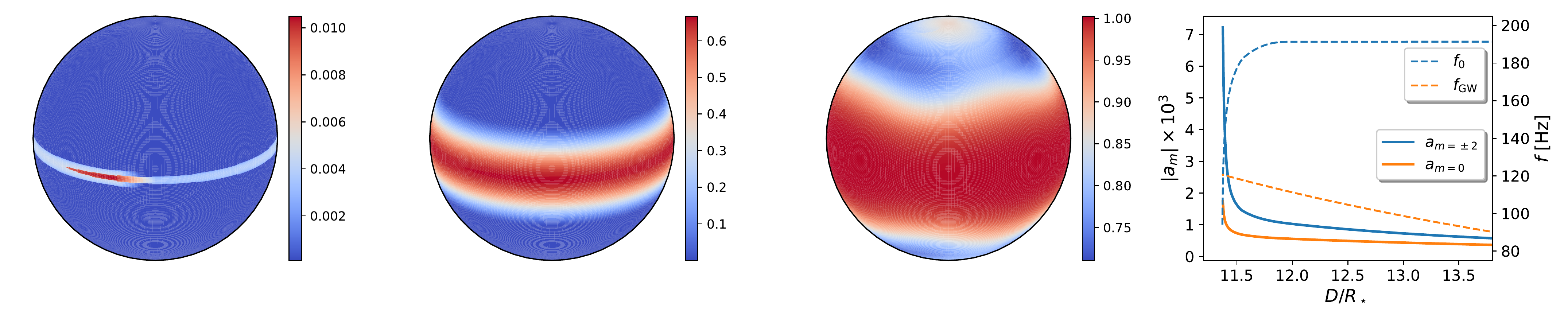}
\caption{\label{fig:melt} Left three panels are the heat maps $e_i/e_{\rm melt}$ of the neutron star crust
(within a $1.3 M_\odot + 1.3 M_\odot$ binary) at binary separations $D = 12.0/ 11.6/11.4 \ R_\star$, respectively.
In the rightmost panel, dashed lines denote the evolution of the i-mode frequency $f_0$ and the gravitational wave (GW) frequency $f_{\rm GW}$,
and solid lines denote the evolution of mode amplitude $a_m$ with $m=\pm 2, 0$.}
\end{figure*}

\vspace{0.2cm}
\noindent{\bf Crust heating up and melting down.} The outer part of the crust is commonly described by a Coulomb lattice with shear modulus $\mu$ \cite{Strohmayer1991}. The inner crust may have nonuniform structures associated with the ``nuclear pasta'' phase \cite{Ravenhall:1983uh,hashimoto1984shape}, which is not considered in this study. Simulations of molecular dynamics \cite{Chugunov2010} have shown that the lattice responds elastically under small applied stress; once the induced strain exceeds the breaking strain ($\epsilon_b \sim 0.1$), plastic deformation starts to develop.
Assuming an applied stress $\sigma$,  the plastic deformation rate $\dot \epsilon_{\rm pl}$ is exponentially small if $\sigma < \sigma_b=\mu \epsilon_b$, and becomes exponentially large if $\sigma > \sigma_b$.
Mathematically, it is well approximated by \cite{Chugunov2010}
\be\label{eq:deform}
\dot \epsilon_{\rm pl} = \frac{n_i Z^2 e^2}{a}\frac{\omega_{\rm p}}{ \mu \bar N \Gamma} e^{(-18.5 \bar {\sigma}_b  +\bar \sigma \bar N) \Gamma}\ ,
\ee
where the dot denotes a time derivative, $\omega_{\rm p}$ is the plasma frequency, $\bar N = 500/(\Gamma-149) + 18.5$, $\bar \sigma = \sigma/(n_i Z^2 e^2/a)$ and $\Gamma= Z^2 e^2/a T$ is the melting parameter with $e$  the electron charge, $Z e$ the total charge per ion, $a$ the lattice spacing, $n_i$ the ion density and $T$ the temperature.
The elastic part of the strain $\epsilon_{\rm el}$  satisfies
$\sigma = \mu\epsilon_{\rm el}$ and the total strain is simply $\epsilon = \epsilon_{\rm el}+\epsilon_{\rm pl}$.

With the plastic deformation, mode energy dissipates into thermal energy, heating up the crust with a rate \cite{Thompson2017}
\be\label{eq:heating}
n_i\dot e_i= \sigma\dot\epsilon_{\rm pl}(\sigma, T)\ ,
\ee
where $n_i$ is the ion number density, $e_i$ is the thermal energy per ion, and $d e_i = c_V dT$ with $c_V$ the specific heat capacity for $T < \Tm$ \cite{Chabrier1993}. Once the melting temperature $\Tm$ is reached, the crustal material still needs an extra amount of latent heat ($\sim k\Tm$ per ion) to be melted \cite{Shapiro1983}. As a result,  the total energy per ion needed to melt the crust from its initial cold state is roughly
$e_{\rm melt} = \int_0^{\Tm} c_V dT + k\Tm$. In this work we have ignored contributions from dripped neutrons as their specific heat may be suppressed by superfluidity.

\vspace{0.2cm}
\noindent{\bf Mode Analysis.}
In the linear approximation, the stellar response to the tidal force is specified by the Lagrangian displacement
$\bm{\xi}(\bm{r},t)$ of a fluid element from its equilibrium position. The displacement can be decomposed into
eigenmodes, $\bm{\xi}(\bm{r},t) = \sum_\alpha a_\alpha(t)\bm{\xi}_\alpha(\bm{r})$,
where $\alpha$ denotes the quantum number of an eigenmode.  In the context of this paper, we only consider
i-modes driven by the leading quadrupole term of the tidal force, so that
$\bm{\xi}_m(\bm{r})= [U(r) \hat r + rV(r)\nabla] Y_{2m}(\theta,\phi)$, where $Y_{2m}(\theta,\phi)$ is the $l=2$ spherical harmonic. The displacement behavior is governed by the  linear pulsation equation \cite{McDermott1985,McDermott1988}
\be\label{eq:L}
[\mathcal L(r;\mu) -\omega_0^2]\bm{\xi}_m=0\ ,
\ee
with $\mathcal L$ being an operator specifying the
restoring force inside the star (see Supplemental Material \cite{Supplemental} for the explicit expression).

For the example star with $M_\star = 1.3 M_\odot, R_\star = 11.7$ km assuming SLy4 EoS \cite{Douchin2001,Read2009} and a core-crust baryon transition density $n_{\rm b,cc} = 0.065\ {\rm fm}^{-3}$,
we obtain an i-mode  frequency $f_0 = \omega_0/2\pi = 190$ Hz \cite{McDermott1988,Tsang2012} and
the tidal coupling coefficient (a measure quantifying the overlap between the waveform and the tidal field)
\be\label{eq:Q}
Q = \frac{1}{M_\star R_\star^2}\int d^3x \rho \ \bm{\xi_m^*}\cdot\nabla[r^2 Y_{2m}(\theta,\phi)] = 0.018,
\ee
 with the normalization $\braket{\xim|\bm \xi_{m'}}:=\int d^3x \rho\ \bm{\xi_m}\cdot\bm{\xi}^*_{m'} = \delta_{mm'} M_\star R_\star^2$, where $\rho$ is the mass density \footnote{In Ref.~\cite{Tsang2012}, a factor $\sqrt{l(l+1)}$ was missed in the normalization calculation.}.

 The evolution of the mode amplitude $a_m(t)$ is governed by \cite{Lai1994}
 \be\label{eq:excitation}
 \ddot a_m + \gamma(t) \dot a_m + \omega_0^2(t) a_m = \frac{GM'W_{2m} Q}{D^3} e^{-im\Phi(t)}\ ,
 \ee
 where the right-hand side is the leading quadrupole term of the tidal driving force with
 $M'=qM_\star$ the companion star mass, $D$  the binary seperation,
 $\Phi(t)$ the orbital phase and $W_{2m}$ is a coefficient of $\mathcal O(1)$ (see Eq.~(2.4) in Ref.~\cite{Lai1994}).
 On the left-hand side,  $\gamma(t)\dot a_m$ is a damping term
 capturing  the plastic deformation induced dissipation with $\gamma(t)$ defined as the ratio
 between the mode energy dissipation rate and two times the mode kinetic energy, i.e.,
 \be\label{eq:gamma}
  \gamma(t)     = \frac{ \int_{\rm crust} n_i \dot e_i  \ d^3x }{M_\star R_\star^2\sum_m | \dot a_m|^2 }\ ,
 \ee
 where the numerator is the crust heating rate (which is equal to the mode energy dissipation rate), and the mode kinetic energy is
 $\frac{1}{2}\int d^3x \rho \bm{\dot\xi}(\bm r,t)\cdot\bm{\dot\xi}^*(\bm r,t)=\frac{1}{2}M_\star R_\star^2\sum_m |\dot a_m|^2$.
 The mode frequency $\omega_0(t)$ to leading order is determined by (see Eq.~(\ref{eq:L}))
 \be\label{eq:omega}
 \omega_0^2(t) = \frac{\braket{\xim|\mathcal L(r; \mu_{\rm avg})\xim} }{\braket{\xim|\xim} }\ ,
 \ee
 where $\mu_{\rm avg}$ is the average shear modulus which decreases as the crust is heated
 and we find the mode frequency is roughtly proportional to the square root of the average shear modulus \cite{Passamonti2012}.

Given the mode amplitude $a_m(t)$, it is straightforward to calculate the fluid element displacement
$\bm{\xi}(\bm{r},t) = \sum_m a_m(t)\bm{\xi}_m(\bm{r})$ and the corresponding strain $\epsilon_{\rm el}$.
From equation~(\ref{eq:deform}), the plastic deformation rate $\dot\epsilon_{\rm pl}$ has an exponential dependence on the
local strain  $\epsilon_{\rm el}$ for $\epsilon_{\rm el} \gtrsim 0.1$, so does the energy dissipation rate
$\sigma\dot\epsilon_{\rm pl}$. Physically, the dissipated energy comes from the local elastic energy, therefore the
energy dissipation rate cannot exceed its replenishment rate $\frac{\mathcal{A}}{2}\mu\epsilon_{\rm el}^2 f_{\rm GW}$,
where $f_{\rm GW}$ is the frequency of both the tidal force and the GW emission and $\mathcal A$ is a coefficient of $\mathcal{O}(1)$.
Here we take $\mathcal A=2$ as an example. As for the initial
condition, we choose $T_i = 0.02\Tm$, where $\Tm\sim 1$ MeV is the melting temperature of the ion crystal at the crust base \cite{Strohmayer1991}. Using the
4th-order Runge-Kutta scheme, we evolve Equations (\ref{eq:deform}, \ref{eq:heating}, \ref{eq:excitation}) on the two-dimensional
surface of the crust base, i.e., we only trace the thermal evolution of the crust base considering its dominant role in the crust heat capacity.

As the neutron star binary spirals inward, the tidal field increases and so does the i-mode amplitude $a_{m=0,\pm2}$, as shown in Fig.~\ref{fig:melt}. At a certain binary separation (with corresponding gravitational wave frequency $f_{\rm GW, melt} < f_0$), part of the crust reaches the yield limit $\epsilon_b$ due to the i-mode excitation and plastic deformation starts.
Heating first takes place at the equator where the strain maximizes.
As the crust heats up, it softens so that
i-mode frequency $f_0$ decreases and the mode amplitude $a_m$ increases. As a result, the crust yields  on larger and larger areas, extending from the equator to the poles, and finally the whole crust is melted. The crust melting takes about $20$
orbit periods and  a total amount of energy $E_{\rm melt}\simeq 1.1\times 10^{47}$ ergs. Notice that this mode treatment is approximate once the plastic motion turns on, where a more accurate description requires 3-dimensional dynamical modeling of crustal motions.
A 2-dimensional consistent evolution was implemented in \cite{Thompson2017} to reveal yield patterns of magnetar crust under strong magnetic stress.

\vspace{0.2cm}
\noindent{\bf Waveform signature.}
After the melting process, part of the binary orbital energy is converted to the mode and thermal energy resulting in a phase shift of the gravitational waveform. Similar to the discussion in \cite{Lai1994,Flanagan2007,Yu:2016ltf} for mode resonances, for the binary neutron star waveform $h(f) = A(f) e^{i \Psi(f)}$, its phase is modified as
\begin{align}\label{eq:dpsi}
\delta \Psi(f) &= \sum_{i=1,2} \delta \phi_i \left ( 1-\frac{f}{f_i}\right ) \Theta(f-f_i)\,\nonumber \\
& \approx \delta \phi_a \left ( 1-\frac{f}{f_a}\right ) \Theta(f-f_a)
\end{align}
where $\Theta$ is the Heaviside function and $f_i$ is the melting frequency of each star. Therefore the search and forecast presented below for crust melting apply equally for generic mode resonances, and we will use `mode resonance signature' and `crust melting signature' interchangeably. The melting process decreases the coalescence phase by $\delta \phi_i$ and the coalescence time by $\delta \phi_i/2\pi f_i$. In the second line we introduced $\delta \phi_a =\sum_i \delta \phi_i$ and $\delta \phi_a/f_a =\sum_i \delta \phi_i/f_i$ to reduce the number of extra parameters in this model, which simplifies the parameter estimation process. Notice that if energy transfers from the orbit to the mode (or heat in this case) during resonance, $\delta \phi$ is positive; if energy transfers from the mode to the orbit, as expected in some of the r-mode resonances \cite{Flanagan:2006sb}, $\delta \phi$ is negative.

\begin{figure}
\includegraphics[scale=0.6]{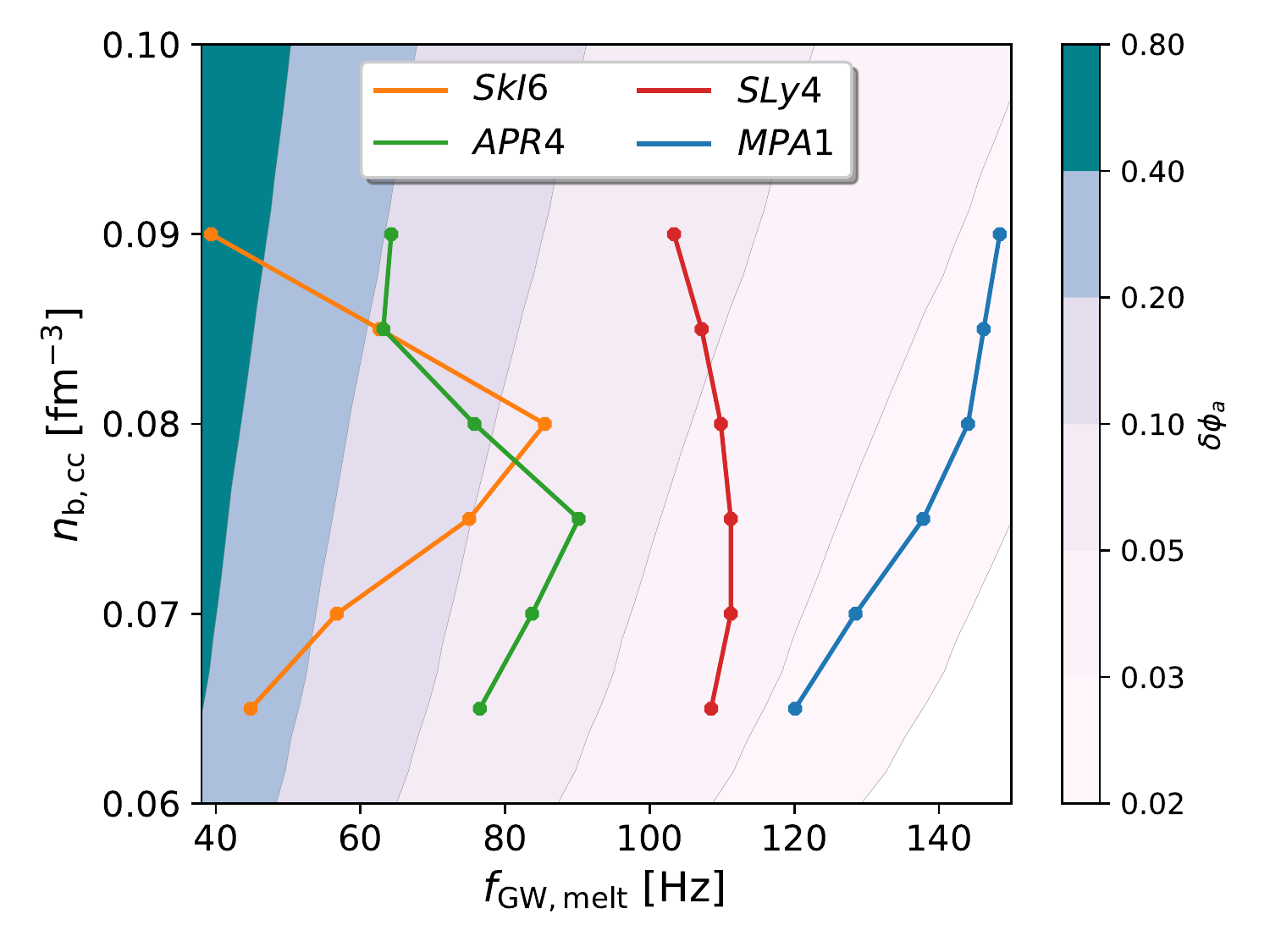}
\caption{\label{fig:phase} The crust melting induced phase change $\delta\phi_a$ in GWs of a BNS merger with each star of $M_\star=1.3 M_\odot$ and $R_\star=12.5/11.3/11.7/12.7$ km for the EoS SkI6/APR4/SLy4/MPA1, respectively, which are not ruled out by the LIGO tidal measurement with GW170817. Note that the core-crust transition density $n_{\rm b,cc}$ is subject
to a large uncertainty in each EoS instead of being an accurately predicted value, so we take the transition density as a free parameter.}\label{fig:f2}
\end{figure}

For each neutron star, $\delta\phi$ depends on its mass $M_\star$, the mass ratio of the companion $q$ (with the companion mass being $qM_\star$), the melting energy $E_{\rm melt}$ and the melting frequency $f_{\rm GW, melt}$ as follows \cite{Lai1994}
\be\label{eq:deltaphi}
  \delta \phi
  =  \frac{2\omega_{\rm orb}E_{\rm melt}}{P_{\rm GW}}
  \simeq  \frac{0.1}{q^2}\left( \frac{1+q}{2}\right)^{2/3}\ E_{47} M_{1.3}^{-10/3} f_{70}^{-7/3}\ ,
\ee
where $\omega_{\rm orb} =\pi f_{\rm GW, melt}$ is the orbital angular frequency, $P_{\rm GW}$ is the energy loss rate due to GW emission, and $E_{47} = E_{\rm melt}/10^{47} {\rm ergs},  M_{1.3} = M_\star/1.3 M_\odot, f_{70} = f_{\rm GW, melt}/70 {\rm Hz}$.
From Equation~(\ref{eq:deltaphi}), we immediately see that the phase shift increases if the melting process happens earlier (lower $ f_{\rm GW, melt}$) in the inspiral phase.
In Fig.~\ref{fig:phase}, we show the total phase change $\delta\phi_a$ for an equal-mass binary neutron star (BNS) merger with  $M_\star = 1.3 M_\odot$,
where $\delta\phi_a$ varies from $0.03$ to $0.6$ depending on the star's EoS and
the core-crust transition baryon density $n_{\rm b,cc}$. The melting energy increases substantially with
increasing $n_{\rm b,cc}$ (commonly assumed to be within $0.06-0.1 \ {\rm fm}^{-3}$ \cite{Horowitz2001,Xu2009,Moustakidis2010}), whereas the i-mode frequency and the associated melting frequency are non-monotonic functions of  $n_{\rm b,cc}$.
We also note that since the mode calculation presented here is Newtonian with the Cowling approximation \cite{Cowling1941},
the fully relativistic mode frequencies may be different (for examples, the frequencies of p- and f-modes
are smaller with the metric perturbation included \cite{Yoshida1997,Chirenti2015}).
If there are also more unpaired neutrons present within the star,
as suggested by the cooling measurement in \cite{Brown2018}, the melting energy may be significantly boosted and the internal mode spectrum
may be modified as well.
Therefore the search of mode resonance signatures may also help probe the superfluid composition of neutron stars.
The effects of nuclear pastas on the melting energy budget and the mode frequency determination also need
to be better understood. Nevertheless, the measurement of $f_a$ and $\delta \phi_a$ will
convey useful information about the core-crust transition density and the star's EoS around that density.

\vspace{0.2cm}

\begin{figure*}
\includegraphics[scale=0.32]{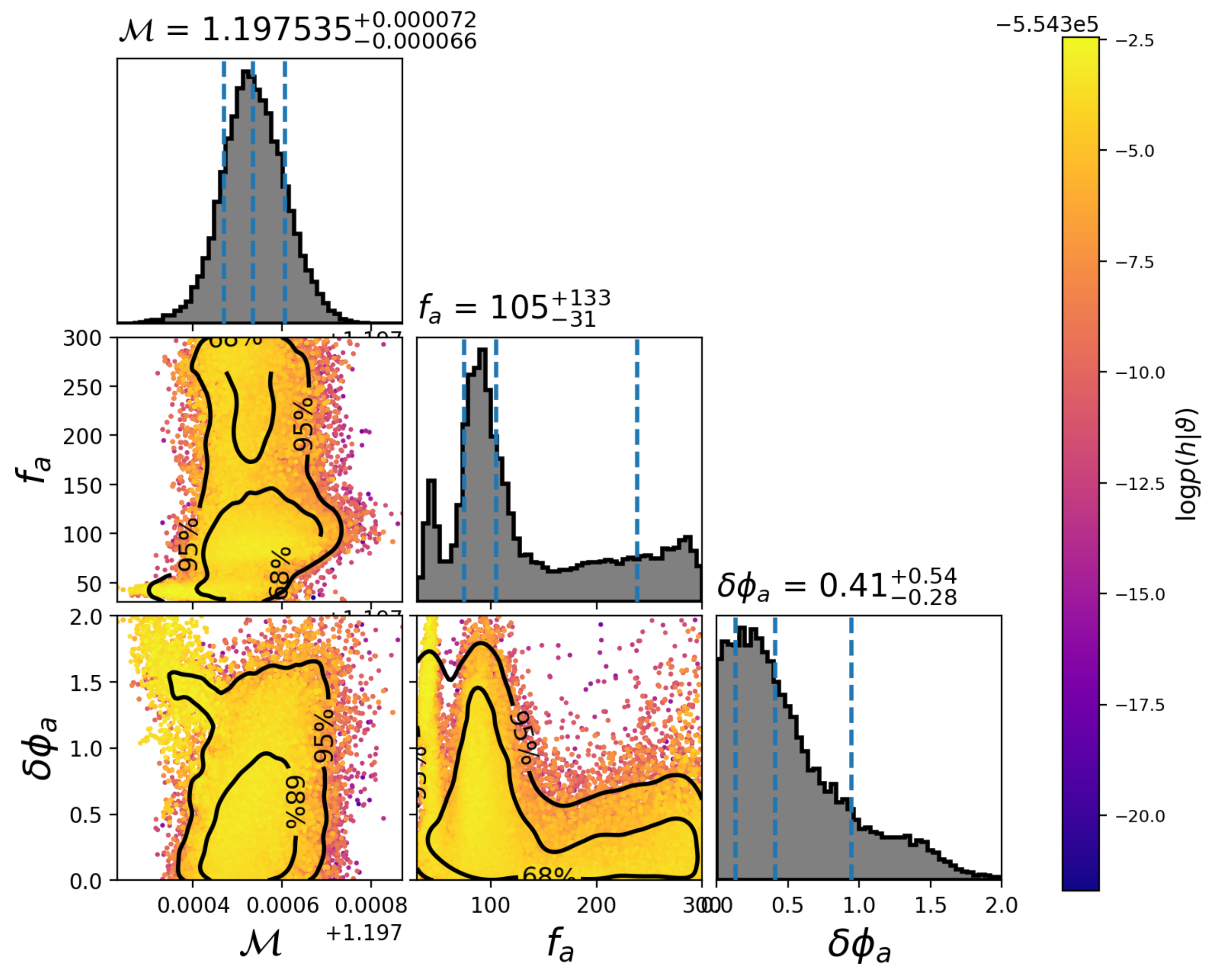}
\includegraphics[scale=0.32]{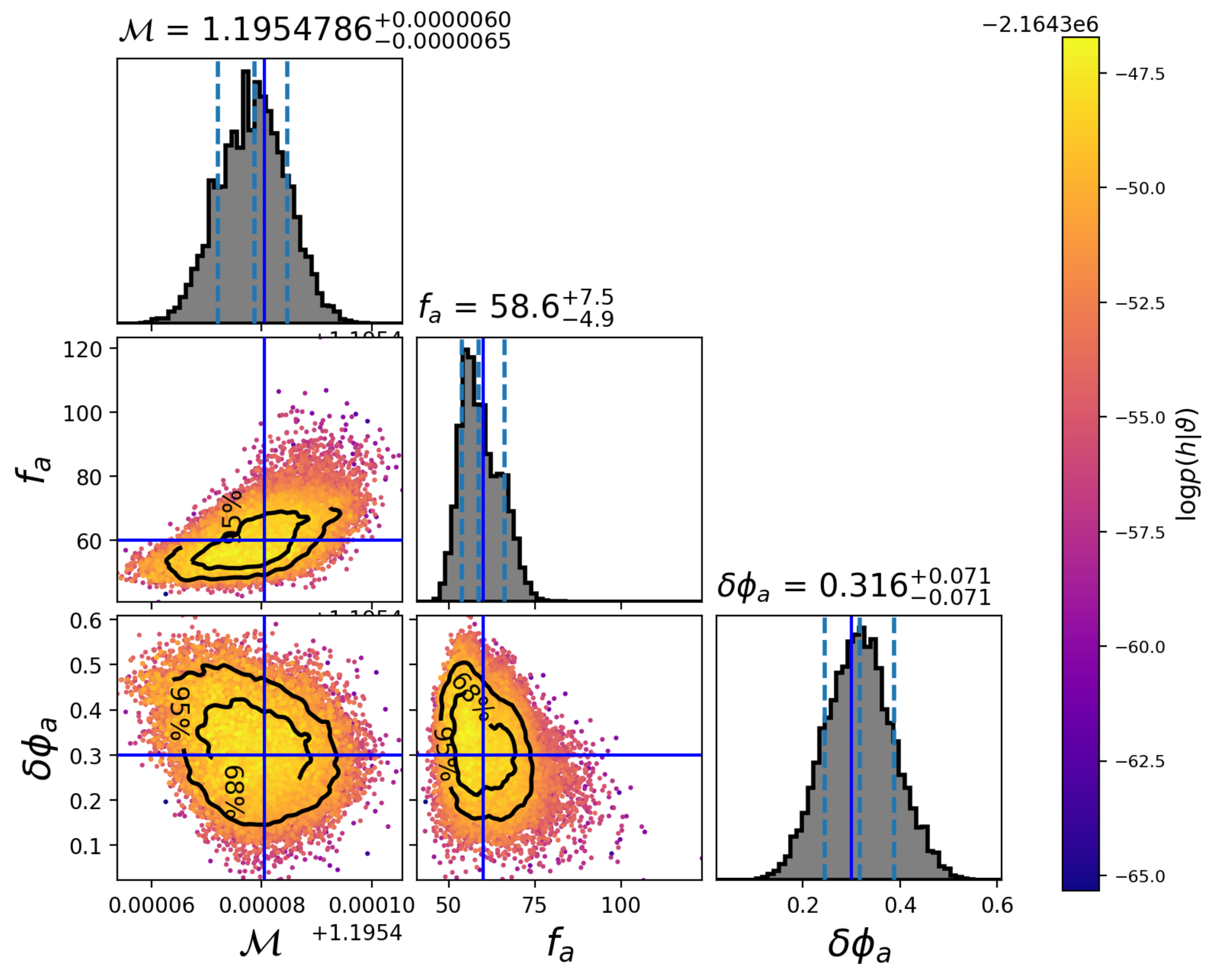}
\caption{\label{fig:p12} Posterior distribution of chirp mass $\mathcal{M}$,  phase shift $\delta \phi_a$ and melting frequency $f_a$ obtained with {\rm PyCBC}, where the prior for $f_a$ is set to be $[30, 300]$Hz and $[0,2]$ for $\delta \phi_a$. Left Plot: the search using data from GW170817. Right Plot: a search obtained assuming LIGO A+ sensitivity and an mode resonance injection at $f_a =60$Hz and $\delta \phi_a =0.3$.}\label{fig:ps}
\end{figure*}

\noindent{\bf Search with GW170817.} We now present the first search for  mode resonance effects (including crust melting) in binary neutron star systems with data from GW170817 with Equation (\ref{eq:dpsi}) implemented. A similar search for tidal-p-g instability is discussed in \cite{Weinberg:2018icl} using different $\delta \Psi(f)$.
The Markov-Chain Monte Carlo (MCMC) parameter estimation is performed with PyCBC \cite{Biwer2019}, for which we assume the source distance and sky location are known as the electromagnetic counterpart of this source has been identified \cite{abbott2017gw170817}. We use the TaylorF2 waveform \cite{Buonanno:2009zt} as the background binary neutron waveform. We present the posterior distributions of chirp mass $\mathcal M$,  $\delta \phi_a$ and $f_a$ in Fig.~\ref{fig:ps}. The marginal distribution of $\phi_a$ indicates that there is no evidence for mode resonance in GW170817, as $\delta \phi_a <1.5$ at $95\%$
confidence level. A similar conclusion can be drawn from a Bayesian model comparison framework. We denote $\mathcal{H}_a$ as the hypothesis with mode resonance and $\mathcal{H}_0$ as the one without, the Bayes factor can be defined as
\begin{align}
\mathcal{B}^a_0 = \frac{P({\rm GW170817} | \mathcal{H}_a )}{P({\rm GW170817} | \mathcal{H}_0 )}
\end{align}
which measures the relative probability of these two hypotheses. We have computed the Bayes factor using both the method of thermodynamic integration \cite{Lartillot2006} and the Savage-Dickey Density Ratio method \cite{Dickey1971}, which both suggest consistent values of $\log  \mathcal{B}^a_0$ in the range of $[-0.6, -0.3]$. This means that these two hypotheses are essentially indistinguishable with this set of gravitational wave data \cite{kass1995bayes}.

It is natural to expect observations  with higher signal-to-noise ratios as the sensitivity of gravitational wave detectors improves.
In the mid-2020s the upgrade of Advanced LIGO, LIGO A+, is expected to start its constrcution \footnote{\url{https://dcc.ligo.org/LIGO-G1601435/public}}. Assuming LIGO A+ design sensitivity for all three detectors at Hanford/Livingston/India,
 and Advanced Virgo  with its full sensitivity, we may observe GW170817-like events with signal-to-noise ratios beyond 100.
In the right panel of Fig.~\ref{fig:ps}, we present a sample search with  an injected signal with $\delta \phi_a =0.3, f_a= 60$ Hz (for a GW170817-type system with star masses $m_1=1.47M_\odot, m_2=1.28M_\odot$, tidal Love numbers $\Lambda_1=210.6, \Lambda_2=525.6$, zero star spins and nearly face-on orientation with inclination angle $\iota = 0.2$ rad) into simulated detector noises consistent with the aforementioned LIGO A+ network sensitivity.
We find the mock signal will be detected with SNR$=197$ and
a MCMC analysis of the mock data successfully recovers the injected values of $f_a$ and $\phi_a$ with small uncertainties.
So it is possible that we observe the crust melting signature in gravitational waves with LIGO A+.

Stacking different events may also improve detectability, as is the case for subdominant modes in black hole ringdowns \cite{Yang:2017zxs}. However, we have no prior information on $\delta \phi_a$ and $f_a$, which are distinct for each binary neutron star system. If we have an underlying or phenomenological model that predicts or characterizes $\delta\phi_a$, $f_a$ as a function of star mass, core-crust transition density and star compactness (which depends on the EoS), the hyper-parameters in this model may be constrained from different events. Certainly the posterior distribution of the hyper-parameters from different events can be multiple together to form the joint probability distribution.
This is something worth to pursue in future studies.

If a mode resonance signature is indeed detected (i.e. preferred over the null hypothesis), it is still necessary to compare to other possible origins, such as tidal-p-g coupling \cite{Weinberg:2018icl}, dynamical scalarization and vectorization \cite{Palenzuela:2013hsa,Annulli:2019fzq}, scalar modes associated to certain GR extensions \cite{Mendes_2018}
and extensions to standard particle physics \cite{Huanag2019}, that predict different $\delta \Psi(f)$. Since the crust melting is nearly instant (Fig.~\ref{fig:melt}),
its impact on the waveform boils down to shifting the coalescence time and the coalescence phase, i.e., $\delta\Psi(f) =\delta\phi_a\times(1-f/f_a)$ with $\delta\phi_a$ being a constant.
For other processes with continuous orbital energy draining,
e.g., the tidal-p-g coupling extending the whole frequency range once the nonlinear instability is turned on,
the waveform signature can be formulated in a similar way except with a frequency dependent phase shift $\delta\phi_a(f)$ which encodes the details of orbital energy draining.
To simulate this, we inject a mode resonance signal ($\delta \phi_a =0.3, f_a=60$Hz) into detector noise corresponding to the LIGO A+ network, and perform the Bayesian model selection between our model resonance waveform and the tidal-p-g waveform.  We find a Bayes factor $\log \mathcal{B}^{a}_{\rm pg} = 2.7\pm 0.3$, suggesting that it is also possible to determine the correct model if a positive detection occurs ( see Supplemental Material \cite{Supplemental} for more details of the Bayesian analysis).
The comparison will be much sharper with third-generation gravitational wave detectors.
 Similarly for the scalarized neutron stars proposed
in scalar tensor theories or other particle physics considerations, there are also effects, such as dipole scalar radiation, that will be effective during the whole frequency range once turned on \cite{Huanag2019}. We also perform a model selection between the mode resonance
and an example model of BNSs with scalar dipole radiation using the same mock data above, and we find the Bayes factor is
$\log \mathcal{B}^{a}_{\rm dipole} = 11.7\pm2.2$ (see Supplemental Material \cite{Supplemental}).

\vspace{0.2cm}
\noindent{\bf Discussion.} Resonant tidal excitations in a neutron star binary induce a phase shift $\delta \phi_a$ in the gravitational wave signal by melting its crust.
In calculating the crust heating rate,
we have used the fitting formula $\dot\epsilon_{\rm pl}(\sigma)$ [Eq.~(\ref{eq:deform})] which is a result of
molecular dynamics simulations \cite{Chugunov2010}. If this simulations result does not accurately apply to the NS crust with a
breaking strain $\epsilon_b$ different from $0.1$, the crust melting frequency $f_{\rm GW, melt}$ will also change. For
a smaller breaking strain $\epsilon_b=0.05$, we find the melting frequency $f_{\rm GW, melt}$ decreases by $\sim 25\%$ and the phase
shift $\delta\phi_a$ increases by a factor $\sim 2$. All the predicted phase shifts corresponding to different EoSs are still well
consistent with the constraint $\delta\phi_a<1.5$ ($95\%$ confidence level) from GW170817.
LIGO A+ may already be able to detect such induced phase shifts.
A 3rd-generation detector network with Cosmic Explorer \cite{Abbott2017CE} sensitivity at the LIGO detectors and Einstein Telescope \cite{Punturo:2010zz} sensitivity at the Virgo detector is able to limit $\delta \phi_a$ with uncertainty $\sim 0.01$ and $f_a$ below 1$\%$. This will not only allow high-confidence detection of the crust melting effect, but also precisely measure crustal and EoS properties as shown in Fig.~\ref{fig:f2}.

We do not expect significant energy release to the neutron star magnetosphere associated with crustal failure, as the magnetic fields ($\sim 10^{12}$G) assumed are too weak to efficiently transfer energy by sending out Alfv\'{e}n waves. However, if the star is a magnetar with field $\sim 10^{15}$ G, this emission mechanism may excite star magnetospheres and power precursor gamma-ray bursts \cite{Thompson2017,Ackermann:2009aa,Troja:2010zm}. Interestingly, the recent LIGO observation of a heavy neutron-star binary (GW190425 \cite{Abbott:2020uma}) may indicate the existence of a fast-merging channel to form binary neutron stars. Such systems may have short-enough lifetime $\sim 10^4$ years to allow active magnetars in the binary coalescence stage \cite{Yang:2020qxt}.

\vspace{0.4cm}
We thank the referees for giving valuable suggestions.
We also thank David Tsang for sharing the code for neutron star mode analysis and
Andrea Passamonti for very helpful discussion. Z. P., Z. L. and H. Y. are supported by the
Natural Sciences and Engineering Research Council of
Canada and in part by Perimeter Institute for Theoretical
Physics. Research at Perimeter Institute is supported in part
by the Government of Canada through the Department of Innovation,
Science and Economic Development Canada and by the Province of
Ontario through the Ministry of Colleges and Universities.

\bibliography{ms}

\newpage
\appendix

\section{Pulsation Equations}
The motion of a mass element inside a star is governed by the continuity equation, the momentum equation and the Possion equation
\begin{equation}\label{eq:pulsation}
  \begin{aligned}
    \frac{\partial\rho}{\partial t} + \nabla\cdot (\rho \bm v) = 0\ , \\
    \frac{\partial \bm v}{\partial t} + (\bm v\cdot\nabla)\bm v = \frac{1}{\rho}\nabla\cdot \bm\sigma -\nabla\Phi\ ,\\
    \nabla^2 \Phi = 4\pi G\rho \ ,
  \end{aligned}
\end{equation}
where $\bm \sigma$ is the stress tensor. In the equilibrium state where $\bm v^{(0)}=0$, the stress tensor is simply
$\sigma_{ij}^{(0)}=-p\delta_{ij}$ with $p$ being the pressure.

The linear pulsation equations can be derived assuming the Lagrangian displacement $\bm\xi(\bm r,t) = \bm\xi(\bm r) e^{i\omega_0 t}$ and the potential perturbation $\delta\Phi(\bm r,t)=\delta\Phi(\bm r)e^{i\omega_0 t}$
with $\{\bm\xi(\bm r), \delta\Phi(\bm r)\}$ and $\omega_0$ being the to-be-determined eigenfunctions and eigenfrequency, respectively.
Consequently, we obtain
$\bm v= \partial\bm\xi(\bm r,t)/\partial t + (\bm v^{(0)}\cdot \nabla) \bm\xi(\bm r,t) = \partial\bm\xi(\bm r,t)/\partial t $
and $\sigma_{ij} = \sigma_{ij}^{(0)} + \Gamma_1 p^{(0)} \epsilon_{kk}\delta_{ij} + 2\mu (\epsilon_{ij}- \frac{1}{3}\epsilon_{kk}\delta_{ij})$, where $\Gamma_1 = (d\ln p/d\ln \rho)_{\rm ad}$ is the adiabatic index,
$\epsilon_{ij}=\frac{1}{2}(\xi_{i,j}+\xi_{j,i})$ is the strain tensor and $\mu$ is the shear modulus.
Plugging them into Eq.~(\ref{eq:pulsation}), we obtain the linear pulsation equation $[\mathcal L(r;\mu)-\omega_0^2]\bm \xi(\bm r)=0$ with \cite{McDermott1985, McDermott1988}
\be\label{eq:lin1}
\begin{aligned}
  \mathcal L(r;\mu) \bm\xi
  &= -\nabla\left(\frac{\Gamma_1 p}{\rho} \nabla\cdot\bm \xi\right) -\nabla\left(\frac{1}{\rho}\bm\xi\cdot\nabla p \right) + \nabla \delta\Phi \\
  & + \frac{1}{\rho}\left[\nabla\left(\frac{2}{3}\mu\nabla\cdot\bm\xi\right)-\left(\nabla\mu\cdot\nabla\right)\bm\xi
    -\nabla(\bm\xi\cdot\nabla\mu) \right] \\
  &  + \frac{1}{\rho}\left[ (\bm\xi\cdot\nabla)\nabla\mu-\mu\left(\nabla^2\bm\xi + \nabla(\nabla\cdot\bm\xi) \right)\right]\ ,
\end{aligned}
\ee
and the linear Possion equation
\be\label{eq:lin2}
\nabla^2 \delta \Phi = -4\pi G (\bm\xi \cdot\nabla\rho + \rho \nabla\cdot\bm\xi)\ .
\ee

For spheroidal modes (for example, the i-mode), the displacement vector can be written as a variable-seperation form
\begin{equation}
  \begin{aligned}
    \bm\xi^r(\bm r) &= U(r)Y_{lm}(\theta,\phi)\ , \\
    \bm\xi^\theta(\bm r) &= V(r)\frac{\partial Y_{lm}}{\partial\theta}(\theta,\phi) \ ,\\
      \bm\xi^\phi(\bm r) &= \frac{V(r)}{\sin\theta}\frac{\partial Y_{lm}}{\partial\phi}(\theta,\phi) \ , \\
      \delta\Phi(\bm r) & = S(r)Y_{lm}(\theta,\phi) \ .
  \end{aligned}
\end{equation}
Plugging them into Eqs.~(\ref{eq:lin1},\ref{eq:lin2}), we obtain the governing equations of $\{U(r), V(r), S(r)\}$ \cite{McDermott1985, McDermott1988}
\begin{equation}
  \begin{aligned}
    \rho\omega_0^2 U=
    & \rho \frac{d\hat\chi}{dr} - \frac{d}{dr}\left(\frac{1}{3}\mu\hat\alpha\right) + \frac{d\mu}{dr}\left(\hat\alpha-2\frac{dU}{dr}\right)\\
    &-\mu\left[ \frac{1}{r^2}\frac{d}{dr}\left(r^2\frac{dU}{dr}\right)-\frac{l(l+1)}{r^2}U + \frac{2l(l+1)}{r^2}V\right]\ , \\
    \rho\omega_0^2 V =
    & \rho\frac{\hat\chi}{r} - \frac{1}{3}\frac{\mu\hat\alpha}{r} -\frac{d\mu}{dr}\left(\frac{dV}{dr}-\frac{V}{r}+\frac{U}{r}\right)\\
    &-\mu\left[\frac{1}{r^2}\frac{d}{dr}\left(r^2\frac{dV}{dr}\right)-\frac{l(l+1)}{r^2}V+\frac{2U}{r^2}\right]\ , \\
    \frac{1}{r^2}\frac{d}{dr}\left(r^2\frac{dS}{dr}\right) &-\frac{l(l+1)}{r^2}S = 4\pi G\left( U\frac{d\rho}{dr}+\hat\alpha\rho\right)\ ,
  \end{aligned}
\end{equation}
with
\[ \begin{aligned}
  \hat\alpha &= \frac{1}{r^2}\frac{d}{dr}(r^2U) - \frac{l(l+1)}{r}V\ ,\\
  \hat\chi   &=-\frac{\Gamma_1 p}{\rho}\hat\alpha - \frac{1}{\rho}\frac{dp}{dr} U + S\ .
\end{aligned}\]
For simplicity, we take the Cowling approximation assuming $\delta\Phi = 0$ and solve the i-mode eigenvalue problem following
Ref.~\cite{McDermott1988}.

\section{\bf Bayesian Parameter Estimation}
\begin{figure*}
\includegraphics[scale=0.75]{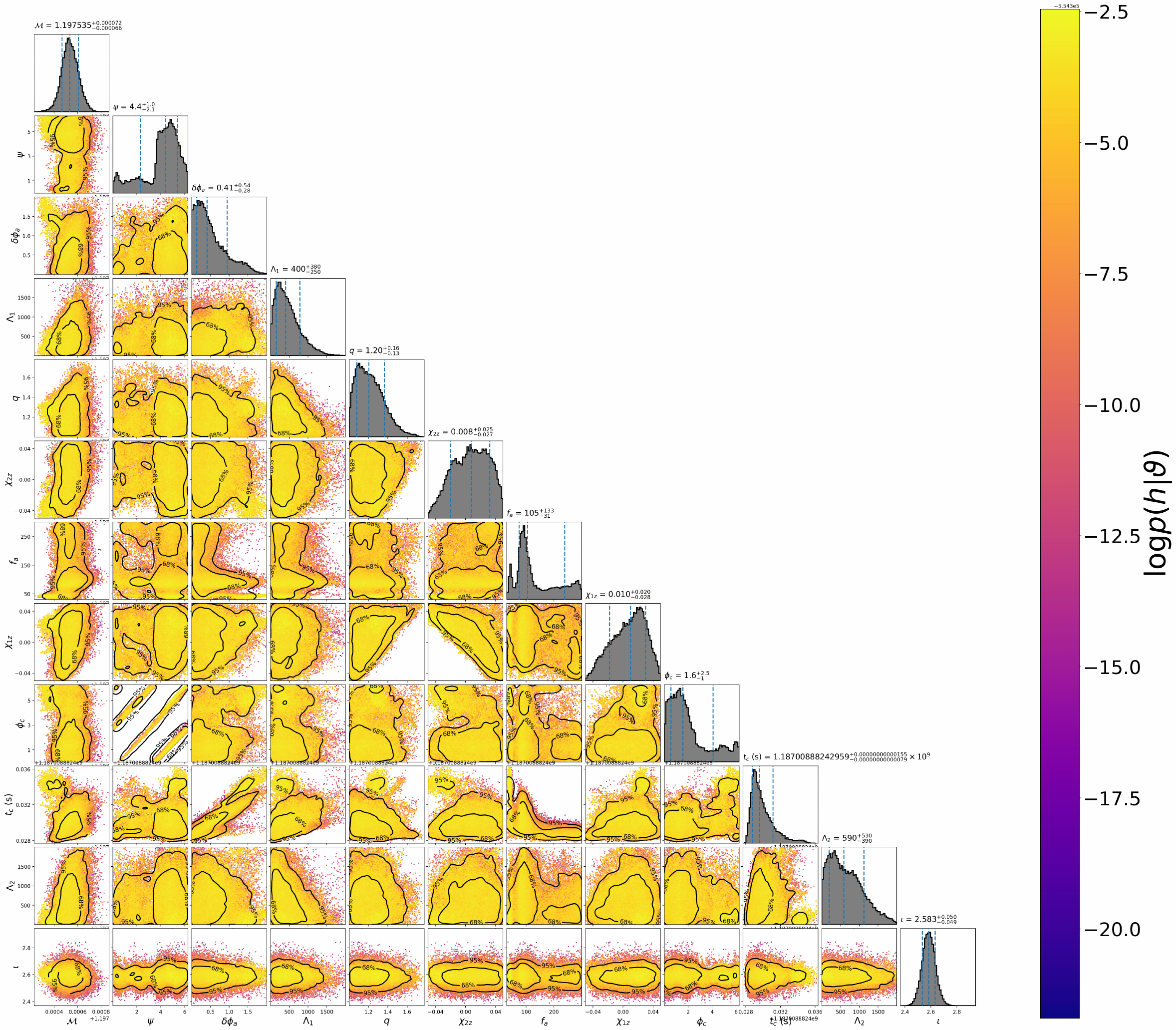}
\caption{\label{fig:fullp} The posterior distribution of all parameters
in the search of mode resonance presented in Fig.~3a in the maintext with data from GW170817.}
\end{figure*}

For the search of possible mode resonance in GW170817, we have incorporated $\delta \phi_a, f_a$ plus all the binary parameters (except for the source distance and sky location which are known from electromagnetic counterparts), including chirp mass $\mathcal{M}$, mass ratio $q$, inclination angle $\iota$, polarization phase $\psi_c$, coalescence phase $\phi_c$, coalescence time $t_c$, tidal Love numbers of both stars $\Lambda_{1,2}$ and parallel spins of both stars $\chi_{1,2 z}$. The priors of the spin are set to be $|\chi_{1,2 z}| <0.05$.
The full posterior distribution of parameters and the Markov-Chain Monte-Carlo samples are presented in Fig.~\ref{fig:fullp}.
In general, the accuracy of the search result not only depends on the event signal-to-noise ratio, but also on the melting frequency. If the melting frequency is too small, even if it is still in the LIGO band, the imbalance of the waveform signal-to-noise ratio before and after the melting process still degrades the search accuracy.
For GW170817, given that the low-frequency sensitivity of the LIGO detectors in O2 is significantly worse than O3, we find that it is beneficial to set the lower bound of the frequency range to be at least 40 Hz to allow ${\rm SNR} \sim 5$ in the waveform before the resonance. This situation will be greatly improved as LIGO reaches design sensitivity when the low-frequency performance is much better, and definitely for LIGO A+ and 3rd-generation detectors, which is important as crust melting may happen before $40$ Hz.

To compare two models or hypotheses, we apply the Bayesian model selection method. For hypothesis $\mathcal{H}_1$ and $\mathcal{H}_0$ and observed data $s$, the Bayes factor is defined as
\begin{align}
\mathcal{B}^1_0 =\frac{P(s | \mathcal{H}_1)}{P(s | \mathcal{H}_0)}\,.
\end{align}
The probability functions ${P(s | \mathcal{H}_{0,1})}$ are usually referred to as the evidence, which may be computed with various tools, such as the thermodynamic integration method \cite{Lartillot2006} and the Savage-Dickey Density Ratio method \cite{Dickey1971}. Larger Bayes factor $\mathcal{B}^1_0$ implies more preference of hypothesis 1 over hypothesis 0, and vice versa.
According to the justification in \cite{kass1995bayes}, if $-1.1 <\log \mathcal{B}^1_0 < 1.1$, the data does not prefer one model over the other; if $1.1 <\log \mathcal{B}^1_0 < 3$, there is positive support for model 1; if $3 <\log \mathcal{B}^1_0 < 5$, there is strong support for model 1 and if $\log \mathcal{B}^1_0 > 5$, the support is overwhelming.
We have applied such formalism in the search for a resonance signature in the data of GW170817, in which case $\mathcal{H}_1$ is the model including the resonance effect and the null hypothesis $\mathcal{H}_0$ is the one without. We obtain $\log \mathcal{B}^1_0 \sim [-0.6, -0.3]$, so that there is no evidence of mode resonance in the parameter range we searched for in the strain data of GW170817.

\begin{figure}
\includegraphics[scale=0.11]{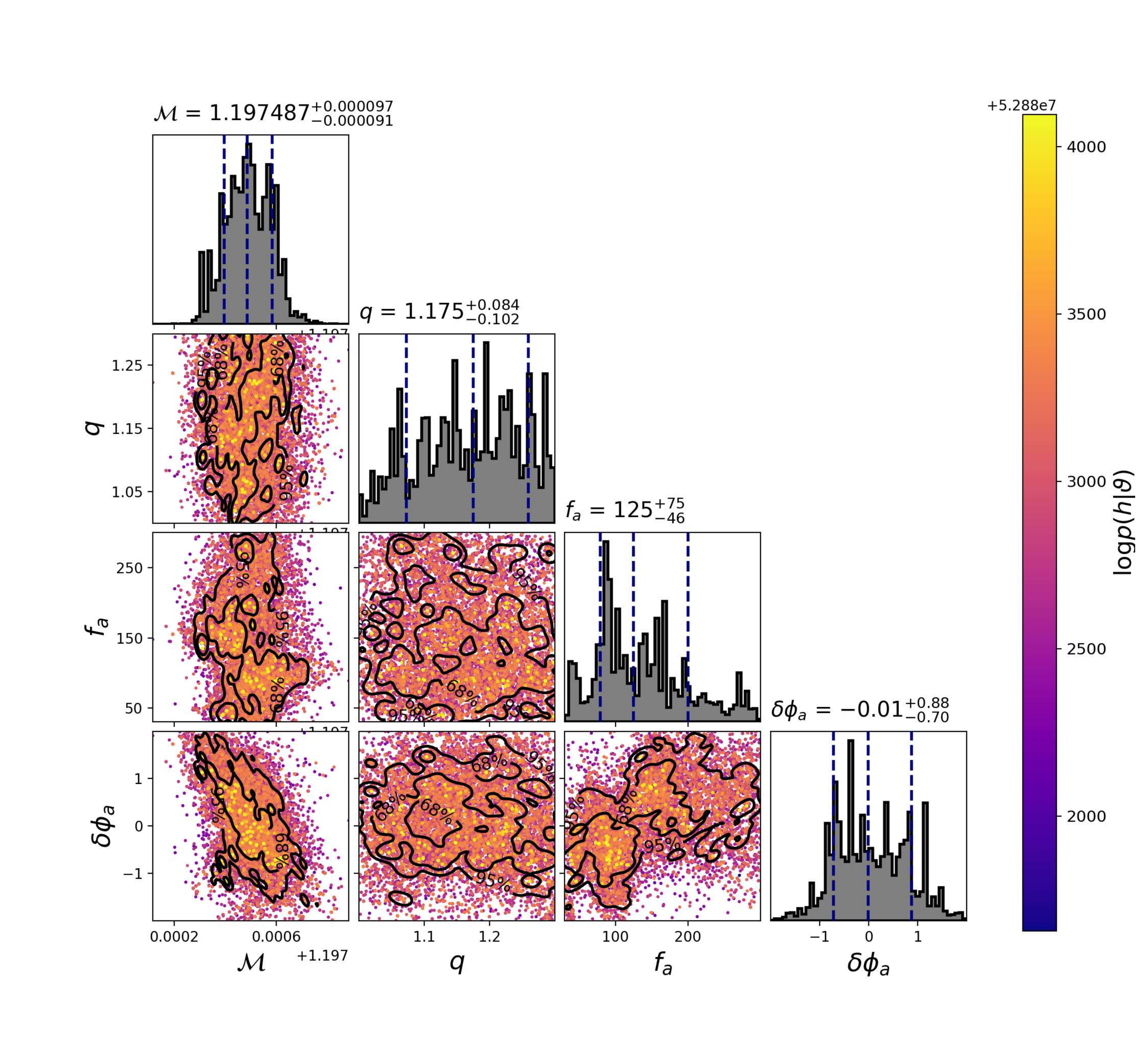}
\caption{\label{fig:wideprior} Posterior distribution in the search of mode resonance from data of GW170817 imposing
a wider prior $[-2,2]$ on $\delta \phi_a$. All the parameter constraints are consistent with the result shown in Fig.~3a in the maintext.}
\end{figure}

For generality, we repeat the above Bayesian analysis imposing a wider prior $[-2,2]$ on $\delta\phi_a$
and a same prior $[30, 300]$ Hz on $f_a$. As a result, we find all the model parameter constraints
are consistent with what shown in the maintext (see Fig.~\ref{fig:wideprior}).

\section{\bf Model selection}

\begin{figure}
\includegraphics[scale=0.33]{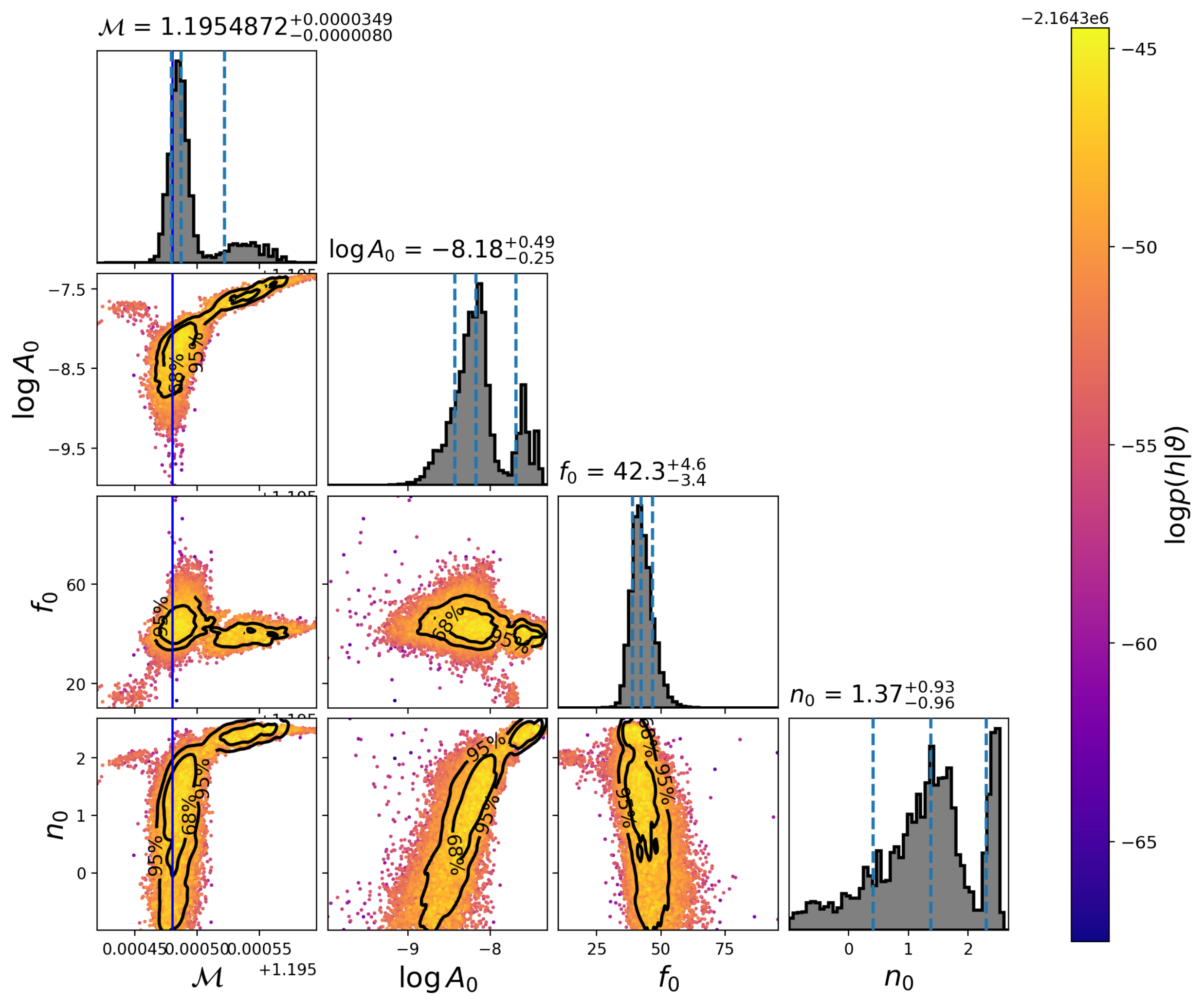}
\caption{\label{fig:pgmode} The posterior distribution of chirp mass $\mathcal{M}$, and $A_0, n_0, f_0$ as we try to fit the mock data with the tidal-p-g mode waveform.}
\end{figure}

The model selection method also applies to distinguish possible origins of the signal. For example, if we detect a signal by searching with our mode resonance waveform, it may also show a positive signal if we had searched for this signal with waveforms motivated by other reasons. To illustrate this, we injected a mode resonance signal ($\delta \phi_a =0.3, f_a=60$Hz) to simulated detector noise compatible with LIGO A+, and searched it with both our mode resonance waveform and the waveform for tidal-p-g coupling \cite{Weinberg:2018icl}:
\begin{align}
\delta \Psi(f) &= -\frac{2C}{3B^2(3-n_0)(4-n_0)} \left \{ \Theta \left( \frac{f}{f_{\rm ref}}\right )^{n_0-3} \right .\nonumber \\
&\left . +(1-\Theta)  \left( \frac{f_0}{f_{\rm ref}}\right )^{n_0-3}  \left[ (4-n_0)-(3-n_0) \left( \frac{f}{f_0}\right )\right ]\right \}\,,
\end{align}
where $\Theta =\Theta(f-f_0)$, $f_{\rm ref} =100$Hz, $C= A_0 [(2m_1)^{2/3}+(2m_2)^{2/3}]/M^{2/3}$, and $B =(32/5) (G \mathcal{M} \pi f_{\rm ref})^{5/3}$.
The corresponding posterior distributions of parameters are shown in ~Fig.~\ref{fig:pgmode}. The fitting with tidal-p-g coupling does not generate a compact posterior distribution of the parameters of this model, $A_0, f_0$ and $n_0$, although the distribution of $\log A_0$ is significantly different from the lower bound of its prior, which is -10.
As we compare the two models, the Bayes factor $\log \mathcal{B}^{\rm res}_{\rm pg}$ is $2.7\pm 0.3$, which shows a preference for the mode resonance model. This means that it is still possible to distinguish these two models when we detect a mode resonance signal with LIGO A+.

\begin{figure}
\includegraphics[scale=0.33]{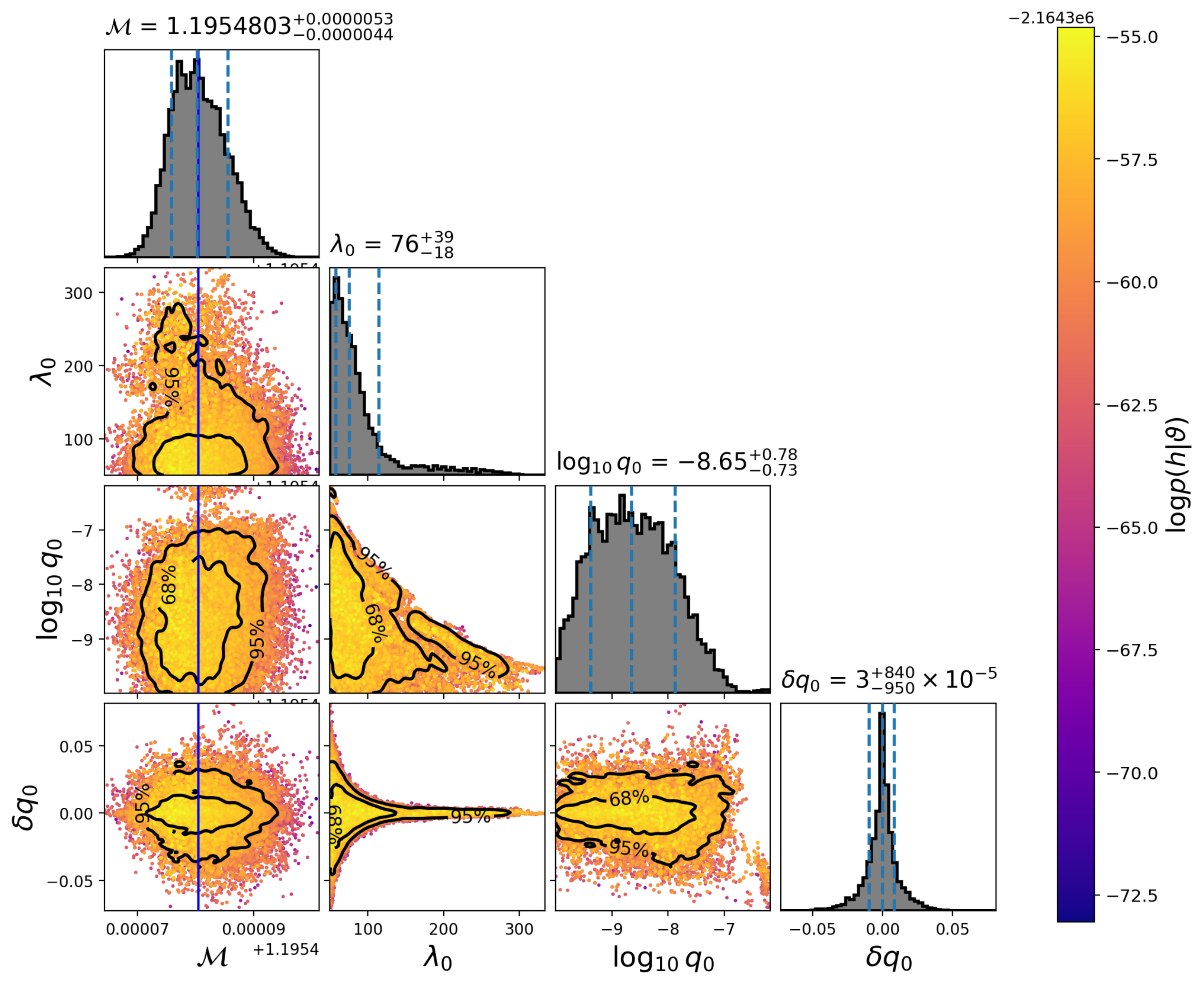}
\caption{\label{fig:dipole} The posterior distribution of chirp mass $\mathcal{M}$,
and $\lambda_0, q_0, \delta q_0$ as we try to fit the mock data with the waveform of scalar dipole radiation.}
\end{figure}

In the case of double NSs carrying scalar (e.g., axions with mass $m_{\rm s}$ \cite{Huanag2019}) charge $q_1$ and $q_2$ , the BNS evolution would be modified by both the extra force mediated by the scalar and the extra scalar dipole radiation.
To the leading order, the extra force can be described in term of Yukuwa potential $V_{\rm s}(r) = -8q_1q_2 e^{-m_{\rm s}r}/r$ and scalar dipole emission power is $P_{\rm s}(r) = \frac{1}{12}\frac{(q_1m_2-q_2m_1)^2}{M^2M^2_{\rm Planck}}\left(1-\frac{m_s^2}{\Omega^2}\right)^{3/2} r^2\Omega^4 \Theta(\Omega^2-m_s^2)$, where $M=m_1+m_2$ is the total mass of the BNS system,
$\Omega(r)$ is the orbital frequency. For convenience, we define symmetry charge $q_0 := q_1q_2/m_1m_2$, anti-symmetry charge
$\delta q_0 :=q_1/m_1-q_2/m_2$ and dimensionless variable $\lambda_0:=(GMm_s)^{-1}$. We find the GW phase shift driven the extra
scalar degree of freedom is $\delta\Psi(f) = 2\pi f\delta t_{\rm s}(f)-\delta\phi_{\rm s}(f)$, with
\be
\begin{aligned}
  \delta\phi_{\rm s}(f) &= \frac{5q_0\lambda_0^{5/2}}{12\eta}
  \Bigg[6\Gamma\left(\frac{5}{2}, \frac{1}{\lambda_0v^2}\right)
  +2\Gamma\left(\frac{7}{2}, \frac{1}{\lambda_0v^2}\right)\\
  &-\Gamma\left(\frac{9}{2}, \frac{1}{\lambda_0v^2}\right) \Bigg]
  +\Bigg[\frac{25(\delta q_0)^2}{1344\eta v^7} \ _2F_1\left(-\frac{1}{2},\frac{7}{6},\frac{13}{6},\frac{1}{\lambda_0^2v^6} \right)\\
  &-\frac{25(\delta q_0)^2}{2496\eta\lambda_0^2 v^{13}} \ _2F_1\left(-\frac{1}{2},\frac{13}{6},\frac{19}{6},\frac{1}{\lambda_0^2v^6}\right)- C_1 \Bigg]\Theta\ ,\\
  \delta t_{\rm s}(f)
  &= \frac{5q_0\lambda_0^4v^3}{24\eta\Omega}
  \Bigg[6\Gamma\left(4, \frac{1}{\lambda_0v^2}\right)
  +2\Gamma\left(5, \frac{1}{\lambda_0v^2}\right)\\
  &-\Gamma\left(6, \frac{1}{\lambda_0v^2}\right) \Bigg]
  +\Bigg[\frac{5(\delta q_0)^2}{768\eta \Omega v^7} \ _2F_1\left(-\frac{1}{2},\frac{5}{3},\frac{8}{3},\frac{1}{\lambda_0^2v^6} \right)\\
  &-\frac{25(\delta q_0)^2}{6144\eta\Omega\lambda_0^2 v^{13}} \ _2F_1\left(-\frac{1}{2},\frac{8}{3},\frac{11}{3},\frac{1}{\lambda_0^2v^6}\right)- C_2\Bigg]\Theta\ ,
\end{aligned}
\ee
where $\Theta:=\Theta(f-\frac{1}{\pi GM\lambda_0})$, $v(f):=(GM\Omega)^{1/3}=(GM\pi f)^{1/3}$, $\Gamma(a,z):=\int_z^\infty t^{a-1}e^{-t} dt$ is the gamma function, $_2F_1$ is the hypergeometric function, and $C_{1,2}$ are two integration constants enabling vanishing $\delta t_{\rm s}$ and $\delta\phi_{\rm s}$
at $f = \frac{1}{\pi GM\lambda_0}$. To illustrate the power of LIGO A+ distinguishing the scalar dipole radiation from the mode resonance, we also constrain the scalar radiation model using the same mock data above  (Fig.~\ref{fig:dipole}) and we find the Bayes factor $\ln \mathcal{B}^{\rm res}_{\rm dipole} = 11.7\pm 2.2$.

\end{document}